\documentstyle[twoside,fleqn,npb,epsfig]{article}
\newcommand{\AmS}{{\protect\the\textfont2
  A\kern-.1667em\lower.5ex\hbox{M}\kern-.125emS}}

\title{
\noindent\hfill\hbox{{\small {\rm  SLAC-PUB-9571}}} \vskip 1pt
NLO Higgs boson rapidity distributions at hadron colliders\thanks{Research supported by the DOE under contract
          DE-AC03-76SF00515 and grant DE-FG02-97ER41022.}}

\author{Charalampos Anastasiou,\address{
          Stanford Linear Accelerator Center,\\ 
          Stanford University, Stanford, CA 94309, U.S.A.}
          Lance Dixon$^{\mbox{{\small \rm a}}}$ 
        and
        Kirill Melnikov\address{Department of Physics and Astronomy,
          University of Hawaii,\\ 2505 Correa Rd. Honolulu, Hawaii 96822,
          U.S.A.}  
        }

\begin{document}
\bibliographystyle{unsrt}

\begin{abstract}
We describe a new method, based on an extension of the unitarity
cutting rules proposed in Ref.~\cite{Anastasiou:2002yz}, 
which is very efficient for the algorithmic evaluation of 
phase-space integrals for various differential distributions.
As a first application, we compute the next-to-leading order normalized
rapidity distribution of the CP-even and the CP-odd Higgs boson
produced in hadron collisions through gluon fusion. We work in the 
heavy top-quark approximation; we find that the NLO corrections at the LHC
are approximately $ 5\% $ in the zero rapidity region.        
\end{abstract}

\maketitle

\section{Introduction}
\label{sec:introduction}

One of the major tasks of 
the Tevatron and the LHC is to discover and explore the  
so far inaccessible Higgs boson sector of the Standard 
Model (SM). The discovery of a single CP conserving Higgs boson, as predicted 
by its minimal version, or a more prolific spectrum of 
Higgs bosons, characteristic to extensions of the SM  such as the
minimal supersymmetric SM (MSSM) or the two-Higgs-doublet-model (2HDM) 
will elucidate the nature of electroweak symmetry breaking.

The dominant mechanism for the production of light Higgs bosons at hadron
colliders is gluon fusion through a heavy quark loop. The production
cross sections of both the CP-even  ($H$) and the CP-odd ($A$)
Higgs bosons are known exactly through 
next-to-leading order (NLO) in perturbative QCD~\cite{higgsnlo} and 
through next-to-next-to-leading order (NNLO) only in the infinite 
top-quark mass approximation~\cite{Anastasiou:2002yz,higgsnnlo}.
Note that, in the case of the CP-odd Higgs boson,  this approximation 
is reliable for small values of $\tan \beta \le 10$, where the
contributions of bottom-quark loops can be ignored. 
The double differential rapidity and $p_T$ distribution for the SM Higgs
boson has been calculated through NLO by means of a fully differential 
Monte-Carlo program\cite{deFlorian:1999zd}, and 
analytically~\cite{nlohiggsdistrib} in the case of
non-zero $p_T$. For the CP-odd Higgs the double differential rapidity and 
$p_T$ distribution was also calculated recently~\cite{Field:2002pb}. 
    
In this paper, we compute the NLO rapidity distributions for the
production of the CP-even and CP-odd Higgs bosons
analytically, including the 
virtual corrections at zero rapidity. This is formally one order lower in
$\alpha_s$ than the NLO contributions of 
Refs.~\cite{deFlorian:1999zd,nlohiggsdistrib,Field:2002pb}. 
(Some numerical results for this distribution, including detector cuts,
have been reported previously~\cite{BDS}.)
For the computation of the inclusive phase-space integrals for fixed Higgs
boson rapidity we extend the method of 
Ref.~\cite{Anastasiou:2002yz} to accommodate the calculation of 
differential distributions.
The idea is to replace the $\delta$-function constraint 
on the phase-space by an ``effective''
propagator. This propagator depends on the constraint and 
in general differs from  conventional particle propagators;
however, if the constraint is polynomial in external momenta, the
resulting Feynman integrals can efficiently be dealt with by algebraic 
means. We will illustrate how this method works in the next Section. 
As a cross-check, we have also computed  the rapidity distributions by
explicitly integrating the finite remainders of the phase-space integrals 
after dipole subtraction~\cite{Catani:1997vz}.  We found complete 
agreement between the two methods.

\section{Method}
\label{sec:method}
For the calculation we use the large top-quark mass 
$(m_t \to  \infty)$ approximation which is known 
to work extremely well~\cite{higgsnlo}, even for relatively large 
Higgs boson masses.
In this limit the interaction of the Higgs boson with gluons
is given by an effective Lagrangian~\cite{higgsnlo,lagrangians} which is 
known to NNLO in the strong coupling constant. 
Keeping only the terms relevant to an NLO calculation, 
the CP-even and CP-odd effective Lagrangians read:
\begin{eqnarray}
\label{eq:Hlagrangian}
&&{\cal L}_{eff}^{H} = -\frac{1}{4v} 
C_1^H \, H \, G_{\mu \nu}^{a} G^{a \mu \nu} \,,
\\
\label{eq:Alagrangian}
&&{\cal L}_{eff}^{A} = \frac{1}{v \tan \beta} 
C_1^A \, A \, G_{\mu \nu}^{a} \tilde G^{a \mu \nu} \,,
\end{eqnarray}
where $G_{\mu \nu}^{a}$ is the gluon strength tensor,
$\tilde G_{\mu \nu} = \epsilon_{\mu \nu \alpha \beta} G^{\alpha \beta}$,
$H,A$ are the Higgs boson fields, and $v \simeq 246$ GeV  is 
the Higgs boson vacuum expectation value.
The Wilson coefficients $C_1^{H,A}$, 
defined in the $\overline{{\rm MS}}$ scheme, 
are~\cite{lagrangians}:
\begin{eqnarray}
&& C_1^H = -\frac{\alpha_s}{3\pi} \Bigg\{ 
 1 + \frac{11}{4} \frac{\alpha_s}{\pi} 
\Bigg\} \,,
\label{eq:hwilson1}
\\
&& C_1^A = - \frac{\alpha_s}{16 \pi} \,,
\end{eqnarray} 
where $\alpha_s = \alpha_s(\mu)$ is the $\overline{{\rm MS}}$ strong coupling
constant defined in the theory with   $n_f=5$ active flavors.

We consider the collision of two hadrons with momenta 
$P_1 = \sqrt{s}/{2} \left (1,{\bf 0}_\perp, 1) \right )$ and 
$P_2 = \sqrt{s}/{2} \left ( 1,{\bf 0}_\perp,-1\right )$, 
producing a Higgs boson $h=\{H,A\}$ with momentum 
$P_h = (E, {\bf p_T}, p_z)$. 
The rapidity $Y$ is defined by:
\begin{equation}
\label{eq:rapidity_definition}
Y = \frac{1}{2} \log \left( \frac{E+p_z}{E-p_z} \right).
\end{equation}
The hadronic rapidity distribution is obtained from the partonic 
rapidity  distributions by convoluting them 
with  appropriate parton densities:
\begin{equation}
\label{eq:sigma_hadronic}
\frac{d \sigma^{h}}{ d Y} = \sum_{i,j} \int_0^1 dx_1 dx_2 f_i(x_1) f_j(x_2)
\frac{d \sigma^{h}_{ij}}{d Y}.
\end{equation}  
The partonic rapidity distributions for the hard scattering of partons 
$i,j$ with momenta $p_1=x_1 P_1$ and $p_2=x_2 P_2$ respectively, are obtained 
by integrating the hard scattering matrix elements over the phase-space of 
the final-state particles with the rapidity of the Higgs boson kept fixed:
\begin{equation}
\label{eq:partonic_distribution}
\frac{d \sigma^h_{ij}}{2 e^{2Y} dY} =  \int d \Pi_f 
\left| {\cal M}^h_{ij}\right|^2 
\delta \left(e^{2Y} - \frac{E+p_z}{E-p_z} \right) \,.  
\end{equation} 
In the center of mass frame of the colliding  hadrons, 
the rapidity constraint can be written as:
\begin{equation}
\label{eq:rapidity_delta}
\delta \left(e^{2Y} - \frac{E+p_z}{E+p_z} \right) 
= e^{-2Y} \delta \left( \frac{P_h \cdot \left[ p_1-u p_2 \right]}
{P_h \cdot p_1} \right) 
\end{equation}
with $u=\left( x_1/x_2 \right) e^{-2Y}$. 

At leading order in $\alpha_s$ a sole Higgs boson is produced;
in this case momentum conservation 
renders the  phase-space integrals trivial.
At NLO, the production of the Higgs boson is accompanied by a production 
of either a quark or a gluon. This makes the phase-space integrations more 
complicated, but they are still sufficiently simple to be done 
directly.  (See for example Ref.~\cite{Altarelli:1979ub} for the
analogous case of Drell-Yan production.)
However, the brute force approach becomes cumbersome at NNLO 
and beyond. 

In this paper we show how to  compute the NLO contributions 
using a method suitable for an algorithmic evaluation of the rapidity 
distribution at NNLO as well. 
The idea is to replace the $\delta$-function constraint 
on the phase space in Eq.~(\ref{eq:rapidity_delta}) 
in terms of the imaginary part of an effective ``propagator'':
\begin{equation}
\label{eq:cutting}
 \delta(x) \to \frac{1}{2\pi i} 
\left [ \frac{1}{x-i0} - \frac{1}{x+i0} \right ].
\end{equation}  
We can then map the constrained phase space integrals onto 
loop integrals in a manner similar to what was suggested for 
unconstrained phase-space integrals in Ref.~\cite{Anastasiou:2002yz}.
It is important that the  constraint in Eq.~(\ref{eq:rapidity_delta})
is a polynomial in momenta; this property allows the application of 
multi-loop algebraic techniques, such as integration-by-parts
and recurrence relations~\cite{IBP}, to the integrals produced after 
the mapping~(\ref{eq:cutting}).

At NLO, using Eqs.~(\ref{eq:rapidity_delta},\ref{eq:cutting}), we 
can express  all the phase-space integrals 
through linear combinations of the  following loop integrals:
\begin{equation}
I(\nu_1, \ldots, \nu_5) = \int \frac{d^d k}{(2\pi)^d} \frac{1}
{A_1^{\nu_1} \cdots A_5^{\nu_5}},
\end{equation}
where
\begin{eqnarray}
A_1 &=& k^2-m_h^2 \pm i\delta, \\  
A_2 &=& (k+p_1)^2, \\
A_3 &=& (k+p_1+p_2)^2 \pm i\delta, \\ 
A_4 &=& (k+p_2)^2, \\
A_5 &=& k\cdot p_1 - uk\cdot p_2 \pm i \delta.
\end{eqnarray}
The propagators $A_1$, $A_3$ and $A_5$ are ``cut'' according to   
Eq.~(\ref{eq:cutting}). 

We now proceed to the reduction of the integrals of the above topology. 
It turns out that we can derive a sufficient set of recurrence relations
through partial fractioning. Integration by parts is not needed for this 
calculation but it will be an essential tool at NNLO.
  
The five propagators of the topology are linearly dependent:
\begin{eqnarray*}
\lefteqn{A_1+A_3-A_2-A_4 = \hat{s} - m_h^2,} && \\
\lefteqn{2A_5+A_1(1-u)+uA_4-A_2 = (u-1) m_h^2,} && 
\end{eqnarray*}
where $\hat{s} = (p_1+p_2)^2$.
Using  the above relations we can eliminate both  propagators 
$A_2$ and $A_4$. It should be noted that 
partial fractioning produces terms with one or more of the cut 
propagators $A_1, A_3, A_5$ eliminated too. These terms have 
a zero contribution to the rapidity constrained phase-space 
integrals and we discard them.
Finally, $I(\nu_1, \ldots, \nu_5)$ reduces to a single  master integral
$I(1,0,1,0,1) = X_1$ in an {\it  algebraic}
fashion. Upon reinstating the $\delta$
functions from the cut propagators, the master integral becomes 
the two-particle phase-space integral evaluated at 
fixed rapidity:
\begin{eqnarray}
\label{eq:master}
\lefteqn{X_1 \to \int  [ {\rm d} h] [{\rm d} q]
~(2\pi)^d \delta^{(d)}(p_1+p_2 - h - q) }
&& \nonumber \\
\lefteqn{ \hskip3cm \times \delta(h \cdot \left[ p_1-u p_2\right ])}
&& \nonumber \\ 
\lefteqn{
 = \frac{
\left[ 
y(1-y)(1-z)^2
\right]^{-\epsilon}
}{(4\pi)^{1-\epsilon}~\hat{s}^{1+\epsilon}~\Gamma(1-\epsilon)} 
\frac{1-y+yz}{1+z} \,, }  &&
\end{eqnarray} 
 where $d=4-2\epsilon$, $[ {\rm d} h] = {\rm d}^{d-1}h/(2h_0 (2\pi)^{d-1})$
 and we have defined:
\begin{equation}
z = \frac{m_h^2}{\hat{s}}, \quad y = \frac{u-z}{(1-z)(1+u)}.
\end{equation}

The real radiation graphs are singular at $z=1$ and $y=0$ or $y=1$. We
extract the poles in $\epsilon$ using identities of the form:
\begin{equation}
y^{-1+\epsilon}= \frac{\delta(y)}{\epsilon} + \sum_{i=0}^{\infty} \left[
  \frac{\log(y)^n}{y}\right]_+ \frac{\epsilon^n}{n!}
\end{equation}
Upon combining the  real and virtual
contributions and  performing the UV renormalization and
mass factorization in
the $\overline{{\rm MS}}$ scheme, all the poles in $\epsilon$ cancel and 
a finite result is obtained for the rapidity distribution. 

\section{Partonic distributions}
\label{sec:part-distr}
We now present the analytic expressions for the partonic rapidity 
distributions.  We write: 
\begin{equation}
\frac{d\sigma^h_{ij}}{d Y}= \frac{ 2 u (1+z)\sigma^h_0}{(1-z) (1+u)^2} 
\frac{d\hat\sigma^h_{ij}}{d y},
\end{equation}
with
\begin{equation}
\frac{d\hat\sigma^h_{ij}}{d y}=
\omega^{h,(0)}_{ij} 
+ \left(\frac{\alpha_s}{\pi}\right) \omega^{h,(1)}_{ij}  
+ {\cal O}\left( \alpha_s^2\right),
\end{equation}
and
\begin{equation}
\sigma^H_0 = \frac{\pi}{576 v^2} \left( \frac{\alpha_s}{\pi}\right)^2,
\quad 
\sigma^A_0 = \frac{9}{4\tan^2 \beta} \sigma^H_0.
\end{equation}
At leading order only the gluon-gluon production channel contributes:
\begin{equation}
\omega^{\{H,A\},(0)}_{gg} = \frac{1}{2} \delta(1-z)
\, \delta(y[1-y]).
\end{equation}
At NLO we obtain contributions from the quark-antiquark, quark-gluon and 
gluon-gluon channels: 
\begin{equation}
\omega^{\{H,A\},(1)}_{q\bar q} =
\frac{16}{9}(1-z)^3 \left[y^2+(1-y)^2 \right],
\end{equation} 
\begin{eqnarray}
\lefteqn{\frac{1}{2}\omega^{\{H,A\},(1)}_{qg}+\frac{1}{2}
\omega^{\{H,A\},(1)}_{gq} 
= -(1-z)^2} && \nonumber \\ \lefteqn{
-\frac{1}{3} \delta(y[1-y]) \Bigg\{\log\left( \frac{z}{(1-z)^2}\right)
\left[1+(1-z)^2\right] } && \nonumber \\
\lefteqn{-z^2 \Bigg\} + \frac{1}{3} \left[1+(1-z)^2\right] \left[ 
\frac{1}{y(1-y)}\right]_+,} &&  
\end{eqnarray} 
\begin{eqnarray}
\lefteqn{
\omega^{H,(1)}_{gg} =  \frac{\delta\left(y[1-y]\right)}{2}
\Bigg\{
\left( 6 \zeta_2 + \frac{11}{2}\right) \delta(1-z) }
&& \nonumber \\ \lefteqn{
 +12 \left[ \frac{\log(1-z)}{1-z}\right]_+ 
-6\left( z^2-z+1\right)^2 \frac{\log(z)}{1-z} }&& \nonumber \\
 \lefteqn { -12z\left(z^2-z+2\right)\log(1-z) \Bigg\} }
&& \nonumber \\
\lefteqn{
+ 
3 \Bigg\{
\left[\frac{1}{1-z} \right]_+ -z\left( z^2-z+2\right)
\Bigg\} \left[ \frac{1}{y(1-y)}\right]_+ }&& \nonumber \\
\label{eq:gghchannel}
 \lefteqn{ -3 (1-z)^3\left[2-y(1-y)\right],} &&
\end{eqnarray} 
\begin{equation}
\omega^{A,(1)}_{gg} =\omega^{H,(1)}_{gg}+\frac{1}{4}\delta(1-z)
 \, \delta(y[1-y]).
\label{eq:AHdiff}
\end{equation} 
The above expressions are valid when the renormalization and factorization 
scales are set  equal to the mass of the produced Higgs boson.  
The full dependence of
the partonic cross sections on those scales can easily be restored by 
solving the renormalization group and DGLAP equations using the above 
expressions as boundary conditions. 

As expected, by integrating the partonic distributions over the rapidity $Y$,  
we obtain the partonic total cross sections calculated 
earlier~\cite{higgsnlo}.

\section{Numerical Results}
\label{sec:numerical-results}

\begin{figure}[h]
\begin{center}
\psfig{figure=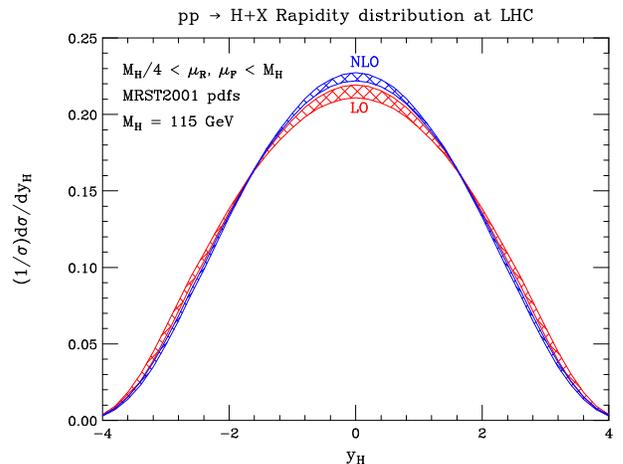,width=80mm}
\end{center}
\vspace*{-1.5cm}
\caption{CP even Higgs boson rapidity distribution at the LHC at 
leading (red) and  next-to-leading order (blue) in perturbative QCD. }
\label{fig:LHC}
\end{figure}

\begin{figure}[t]
\begin{center}
\psfig{figure=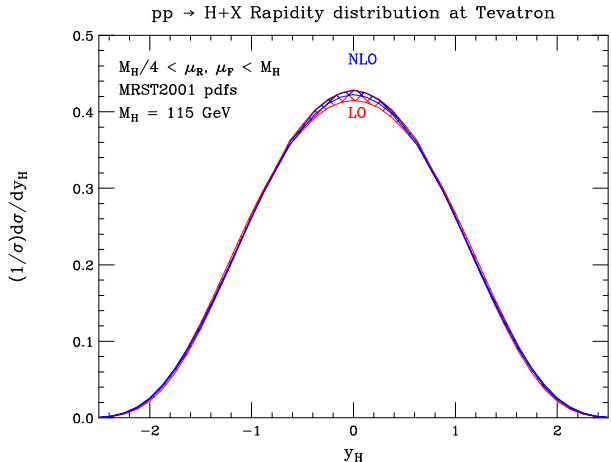,width=80mm}
\end{center}
\vspace*{-1.5cm}
\caption{CP even Higgs boson rapidity distribution at the Tevatron at 
leading (red) and  next-to-leading order (blue) in perturbative QCD. }
\label{fig:Tev}
\end{figure}

In this section we present numerical results for the NLO  
rapidity distributions of the CP-even and CP-odd Higgs bosons at the LHC
and the Tevatron. 
We calculate the hadronic rapidity distributions by convolving
the partonic cross sections of the previous section with the NLO parton
distribution functions, as in Eq.~(\ref{eq:sigma_hadronic}). 
The resulting rapidity distributions, normalized to the 
total cross section, are shown in Fig.~\ref{fig:LHC}. 
Fig.~\ref{fig:Tev} shows the analogous plots for the Tevatron.
It is clear from the plots that the corrections to the shape of the 
distributions are fairly small.  For example, at zero rapidity, where 
the corrections are largest, they increase the LHC result by 
only $5\%$. For the Tevatron, the NLO rapidity distribution falls within 
the band of the LO distribution. The dependence on the factorization
and renormalization scales is also small, suggesting that 
higher order perturbative corrections are unlikely to be large.

This very stable behavior should be contrasted with the known 
fact that the NLO corrections to the total Higgs boson 
hadroproduction cross section are very large; they 
increase the cross section by approximately a factor $1.7$.
Our result indicates  that since the shape of the distribution 
is very stable against higher order QCD corrections, it 
can be reliably  predicted even by LO Monte Carlo event generators 
normalized  to the NNLO  results for the total cross section. This procedure
should give a fairly accurate description of the Higgs 
rapidity distribution at the LHC.

In the case of the CP-odd Higgs boson, the situation is rather similar.
The partonic cross sections 
for the CP-even and the CP-odd Higgs bosons differ only in a single 
term proportional to $\delta(1-z)$ with a small coefficient, 
Eq.~(\ref{eq:AHdiff}); 
therefore the  rapidity distributions 
for the CP-odd Higgs boson are numerically very  similar to the 
distributions shown in Fig.\ref{fig:LHC} and Fig.\ref{fig:Tev}.

\section{Summary}
\label{sec:conclusions}

In this paper we computed the NLO rapidity distribution of CP-even 
and CP-odd Higgs bosons produced at hadron colliders.   We found that 
the NLO corrections change the rapidity distribution, normalized to the total
hadronic cross section, only by a small amount. 
For example, at zero rapidity the NLO normalized distribution for the LHC
increases by approximately $5\%$ as compared to LO. 
The scale variation decreases by a factor of two, from LO to NLO. 

The phase-space integrations of the real radiation graphs 
with fixed rapidity of the Higgs boson are straightforward at 
this order in perturbation theory. However, traditional methods are 
very cumbersome for the evaluation of the Higgs boson rapidity distributions 
at NNLO. In this paper, we performed the first test of the 
extension of the method suggested in Ref.~\cite{Anastasiou:2002yz} 
for evaluating phase-space integrals using multiloop techniques, 
by applying it to the rapidity distribution of a hadroproduced
color-singlet state.

We are confident that the same method is tractable for the evaluation of 
the differential distributions in more complicated  cases, 
such as the rapidity distribution for Drell-Yan and
Higgs boson hadroproduction at NNLO. 
This will be the subject of a future work.


\end{document}